\documentclass[final,3p]{elsarticle}

\usepackage[utf8]{inputenc}
\usepackage[T2A]{fontenc}
\usepackage[english]{babel}
\usepackage{amssymb}
\usepackage{amsmath}
\usepackage{hyperref}
\usepackage{color}
\usepackage{soul}

\newcommand{\E}{{\bf E}}
\newcommand{\B}{{\bf B}}
\newcommand{\xp}{{\bf x}_p}
\newcommand{\vp}{{\bf v}_p}
\newcommand{\rot}{\mbox{\rm{rot\,}}}
\newcommand{\diverg}{\mbox{\rm{div\,}}}

\journal{Computer Physics Communications}

\begin{document}

\begin{frontmatter}

\title{A fully implicit electromagnetic energy- and charge-conserving PIC model for simulations of high-$\beta$ plasma \tnoteref{t1}}

\tnotetext[t1]{The work is supported by Russian Science Foundation (grant \textnumero~25-11-00096)}

\author[nsu,binp]{V.A. Kurshakov\corref{cor1}}

\ead{V.A.Kurshakov@inp.nsk.su}

\author[nsu,binp]{I.V. Timofeev}

\affiliation[nsu]{
    organization={Novosibirsk State University},
    addressline={Pirogov st., 1}, 
    city={Novosibirsk},
    postcode={630090}, 
    country={Russian Federation}
}

\affiliation[binp]{
    organization={Budker Institute of Nuclear Physics SB RAS},
    addressline={Lavrent'ev av., 11}, 
    city={Novosibirsk},
    postcode={630090}, 
    country={Russian Federation}
}

\cortext[cor1]{Corresponding author}

\begin{abstract}
The paper presents a 3D numerical model of plasma in which a fully implicit
PIC method with exact conservation of global energy and local charge is
generalized to the electromagnetic case. This opens the possibility of
simulating plasma equilibria with a high (up to an extreme) ratio of plasma
pressure to magnetic field pressure, which are formed in open magnetic traps.
The main goal of this work is to find out whether a fully implicit model,
which usually performs numerous nonlinear iterations when
solving the coupled Vlasov--Maxwell system, can compete in performance with
semi-implicit models that achieve the same conservative properties at the cost
of only a single particle push per time step. It turns out that in 3D
geometry such competition is quite possible due to the growing computational
cost of evaluating the mass matrices in the semi-implicit approach. The
software implementation of the proposed numerical model is written in C++
within the parallel \texttt{xpic} code. Testing the code on problems of the
development of the Weibel instability and the formation of plasma with
extreme pressure showed good agreement between the semi-implicit and fully
implicit approaches and made it possible to establish their relative speed.
\end{abstract}


\end{frontmatter}

\section*{Introduction}

In the physics of plasma confinement in open magnetic traps, of
particular interest is the class of regimes with an extremely high ratio of
plasma pressure to magnetic pressure, the parameter $\beta = 8\pi P / B_v^2$
approaching unity. The formation of a compact plasmoid whose
pressure $P$ becomes comparable to the pressure of the external vacuum magnetic
field $B_v^2/(8\pi)$ opens new possibilities for plasma confinement. In such
regimes, the diamagnetic effect leads to partial or complete expulsion of the
magnetic field from the plasma volume, which substantially changes the topology
of the field lines. As a consequence, suppression of longitudinal losses of
particles and energy through the magnetic mirrors is expected due to the
increase of the effective mirror ratio and the reduced heat conduction
along the magnetic field.

Several scenarios for the realization of such equilibria have been proposed.
In open traps, these are the formation of a field-reversed configuration
(FRC)~\cite{roche2025}, in which the longitudinal field changes sign, or the
formation of a diamagnetic bubble~\cite{beklemishev2016}~--- a localized region
with a strongly weakened field, bordering the vacuum magnetic field through a
thin transition layer. Similar high-$\beta$ regimes with pronounced diamagnetic
effects are also studied in other magnetic confinement concepts, such as the
Polywell~\cite{park2026}. Currently, the CAT facility has been commissioned at
the Budker Institute of Nuclear Physics~\cite{bagryansky2016,kolesnichenko2026},
where a plasmoid with a record $\beta$ is planned to be created by powerful
injection of fast neutral atoms into a pre-created target plasma. This requires
an adequate theoretical and numerical description of the processes both at the
stage of bubble formation and in the regime of its quasi-steady existence.

In the numerical simulation of plasma processes in open systems, the
fundamental problem lies in the natural separation of spatial and temporal
scales. Plasma exhibits a wide spectrum of collective phenomena: from
small-scale kinetic effects that determine the structure of shock waves and
transition layers, to large-scale hydrodynamics associated with macroscopic
stability and transport. For a correct reproduction of high-$\beta$
confinement regimes in open traps, two important physical problems must be
resolved simultaneously. First, this is the physics of the transition layer of
the diamagnetic bubble, where a sharp boundary is formed between regions with
strong and weak fields. The width of this layer is found to be affected not
only by ion, but also by electron kinetics~\cite{kotelnikov2020,park2019,kurshakov2023,timofeev2024}.
Second, these are the processes of electron component transport in the
magnetic expander~--- the region connecting the mirror with the absorbing wall.
Electrons provide the electrical contact of the plasma with the wall and form
near-wall potentials that directly affect the ion flux and the overall energy
balance of the system. These processes require a kinetic description of not
only ions but also electrons, whose dynamics determines the ambipolar fields
and the electron heat conduction.

However, most existing numerical models are limited to hybrid approaches, in
which ions are treated kinetically (e.g., by the particle method), while for
electrons a simplified assumption of a Boltzmann distribution with a uniform
temperature along the field lines is adopted. Such an approximation is
justified in closed systems with high longitudinal heat conduction, but in open
configurations it is violated due to the presence of losses and temperature
gradients. In particular, it does not allow one to correctly describe the
physics of the expander, where electrons are not in thermodynamic equilibrium~\cite{glinskiy2024,glinskiy2026,tyushev2025}.
The alternative path~--- a gyrokinetic description of electrons~--- also turns out
to be inapplicable at high beta ($\beta\approx 1$), since it assumes the
presence of a strong guiding magnetic field and a small Larmor radius compared
to the field gradients, whereas in a plasmoid with extreme $\beta$ regions with
zero or strongly weakened magnetic field are formed, where the Larmor frequencies
tend to zero and the adiabatic invariant ceases to exist.

Thus, the only correct path is the use of fully kinetic models that solve the
Vlasov--Maxwell equations for all plasma components. An acceptable simplification
in this case is the decision not to resolve fast oscillations at the electron plasma
frequency $\omega_{pe}$ and spatial scales of the order of the Debye radius
$\lambda_D$, since these scales, most of the time, are insignificant for the
macroscopic evolution of the confined plasma. However, classical explicit
particle-in-cell (PIC) schemes turn out to be practically inapplicable in such
conditions from the standpoint of performance: the rigid Courant--Friedrichs--Lewy
(CFL) condition at typical parameters of open traps requires a time step much
smaller than the electron cyclotron rotation period, and the grid must resolve
the Debye radius. As a result, simulations over confinement times (of the order
of milliseconds) become excessively resource-consuming. Therefore, a transition
to implicit finite-difference approximations of the particle equations of
motion and Maxwell's equations is necessary, removing the CFL constraint and
allowing large steps in time and space.

Equally critical is the exact conservation of the total energy of the system
(fields and particles): the characteristic confinement times exceed the
electron plasma and cyclotron periods by many orders of magnitude, and any
artificial dissipation or energy pumping can completely distort the transport
pattern. Among energy-conserving approaches, the most widespread are the
semi-implicit ECSIM schemes~\cite{lapenta2017,lapenta2017ms,lapenta2023,ren2024,berendeev2026} and fully implicit
methods~\cite{markidis2011,chen2011,chen2023}. The semi-implicit ECSIM scheme
is attractive because the field equations remain linear and the positions and
velocities of particles are not coupled, which gives high efficiency per
particle. The price of this, however, is the need to assemble and store at each
step the nonlocal Lapenta mass matrix $M_{gg'}$, through which the current is
expressed linearly in terms of the electric field on the grid. In one- and
two-dimensional problems, the overhead of filling this matrix is usually
acceptable, but in three-dimensional geometry the situation deteriorates
sharply: already under the assumption of a single cell crossing per step, the
matrix contains on the order of $2\cdot 10^3$ components per grid node, which
in terms of memory is equivalent to $\sim 150$ computational particles per
cell~\cite{angus2023}. For a moderate number of particles per cell, filling the
Lapenta matrix becomes one of the most time-consuming stages of the computational cycle, and
the advantages of ECSIM in performance noticeably weaken~\cite{angus2023}. In
addition, ECSIM is incompatible with charge-conserving current decomposition,
so to satisfy Gauss's law one has to introduce current correction
~\cite{pinto2022,berendeev2024} or use analogues of Marder
smoothing~\cite{marder1987,nielsen1990,chen2019}.

Fully implicit schemes require the joint solution of the nonlinear
field--particle system (usually by the particle-suppressed JFNK method) and are
therefore more expensive per step, but they fundamentally avoid the mass
matrix: the current is computed directly from the particles, and the continuity
equation is ensured by a consistent interpolation of the electric field along
the trajectory. For three-dimensional simulations of high-$\beta$ plasma, where
the grid volume is large and the number of particles per cell is limited, such
an approach may turn out to be more advantageous in terms of memory and
scalability, even if in two-dimensional tests it is inferior to ECSIM in step
time.

In this work, we follow the approach of Chen et
al.~\cite{chen2023} and formulate a fully implicit electromagnetic PIC scheme
that simultaneously conserves the total energy and the local electric charge.
In contrast to semi-implicit implementations based
on~\cite{lapenta2017,lapenta2023}, the continuity equation is satisfied without
additional charge or current corrections~--- due to a consistent discretization
in which the change of charge in a cell is exactly equal to the flux of current
through its boundaries.

The developed scheme has been verified on the same set of tests as the
semi-implicit model~\cite{berendeev2024}: (i)~a stable equilibrium of plasma
with uniform density; (ii)~the linear stage of the Weibel instability;
(iii)~dynamic expulsion of the magnetic field under continuous plasma
injection~--- a key test of the ability to describe the formation of a
diamagnetic bubble. Particular attention is paid to the comparison of
performance with the semi-implicit scheme. The results confirm the conservative
properties of the model and its suitability for simulating high-$\beta$ plasma.

\section*{Model description}

The goal of this section is to extend the algorithm proposed in
Ref.~\cite{chen2023} to the case of the evolution of both the electric and
magnetic fields, while preserving the main advantages of the algorithm---the
possibility of conserving energy and charge. In contrast to the original
work~\cite{chen2023}, where only the electrostatic problem was considered and
the magnetic field was assumed constant and uniform, here we allow variations
of the magnetic field $\B_g^{n}$ and the electric field $\E_g^n$ on a standard
Yee grid. This yields a self-consistent electromagnetic PIC model in which
particles move under the action of the fields, and the fields, in turn, evolve
under the action of the current collected from the particles. As in
Ref.~\cite{chen2023}, the key requirement remains the consistency of the fields 
interpolation from grid to particles and of the current decomposition from
particles to grid: without it, the discrete conservation laws are violated.

\subsection*{Particle equations of motion}

Let us begin with the description of the equations of motion for a particle $p$
with coordinates $\xp^n$ and velocities $\vp^n$:
\begin{align}
    \xp^{n+1} &= \xp^n + \Delta t \vp^{n+1/2},\label{eq:x}\\
    \vp^{n+1} &= \vp^n + \frac{q_p \Delta t}{m_p} \left( \E_p^{n+1/2} + \left[
    \vp^{n+1/2} \times \B_p^{n+1/2} \right] \right).\label{eq:v}
\end{align}
Here the half-step quantities are defined as $A^{n+1/2} = (A^{n} +
A^{n+1})/2$, $\Delta t$ is the simulation time step, and $q_p$ and $m_p$ are
the charge and mass of particle $p$. In what follows, a dimensionless system of
units is used: charge is measured in units of the elementary electron charge
$e$, mass in electron masses $m_e$, time in $\omega_{pe}^{-1}$, where
$\omega_{pe} = \sqrt{4 \pi e^2 n_0 / m_e}$ is the plasma frequency evaluated
for a characteristic density $n_0$, velocities in units of the speed of light
$c$, distances in $c/\omega_{pe}$, and the electric and magnetic fields in
$m_e c \omega_{pe} / e$. The electric field $\E_p^{n+1/2}$ is interpolated to
the particle from the grid using a kernel averaged along the path between the
points $\xp^{n}$ and $\xp^{n+1}$, which ensures the conservation of energy and
charge. The magnetic field $\B_p^{n+1/2}$, since it does no work on the
particle, can be interpolated in the standard way for the half-step coordinate
$\xp^{n+1/2}$ using a second-order b-spline $S_2$:
\begin{equation}
    \B_p^{n+1/2} = \sum_g \B_g^{n+1/2} S_2(\xp^{n+1/2} - {\bf
    x}_g).\label{eq:bp}
\end{equation}
The difference in the interpolation of the electric and magnetic fields is
related to the work of the Lorentz force: the magnetic field does no work,
therefore its value can be taken at the midpoint of the trajectory
$\xp^{n+1/2}$ without compromising the fulfillment of the energy conservation
law. The electric field, on the contrary, must be averaged along the entire
trajectory of the particle, since only its work changes the kinetic energy. The
use of a standard interpolation at a single point would lead to a nonphysical
energy influx on scales of the order of the grid step.

In Eqs.~(\ref{eq:x}, \ref{eq:v}), the only independent variable that remains is
$\vp^{n+1/2}$, since the coordinate at the new step is always given by
Eq.~(\ref{eq:x}). Equation~(\ref{eq:v}) is nonlinear, because the electric and
magnetic field quantities are determined, among other things, by the coordinate
at the new step $\xp^{n+1}$. To solve it, the Picard iteration method is used;
in practice, this process converges within a few iterations and can operate at
a relatively large step $q_p B_p \Delta t / m_p \gtrsim 1$
\cite{markidis2011,chen2023,koshkarov2022}. At each iteration,
Eq.~(\ref{eq:v}) is inverted analytically:
\begin{align}
    {\bf a}_p &= \vp^{n} + \alpha_p \E_p^{n+1/2, \, k},\nonumber\\
    {\bf b}_p &= \alpha_p \B_p^{n+1/2, \, k},\nonumber\\
    \vp^{n+1/2, \, k+1} &= \frac{{\bf a}_p + {\bf a}_p \times {\bf b}_p + ({\bf
    a}_p \cdot {\bf b}_p) {\bf b}_p}{1 + b_p^2},\label{eq:vh}\\
    \xp^{n+1, \, k+1} &= \xp^{n} + \Delta t \vp^{n+1/2, \, k+1},\nonumber\\
    \vp^{n+1, \, k+1} &= 2 \vp^{n+1/2, \, k+1} - \vp^{n},\nonumber
\end{align}
where $\alpha_p = 0.5 \, q_p \Delta t / m_p$. As the first approximation at step
$k=0$, the quantities from the previous time step $\xp^{n}, \vp^{n}$ are used.
However, in the presence of magnetic field gradients comparable to the Larmor
radius, such particle motion algorithm describes the dynamics incorrectly
~\cite{vu1995,rickeston2020}. The elimination of this limitation will be
considered in future work.

\subsection*{Consistent interpolation of the field and current}

\begin{figure}[htb!]
    \centering
	\includegraphics[scale=0.8]{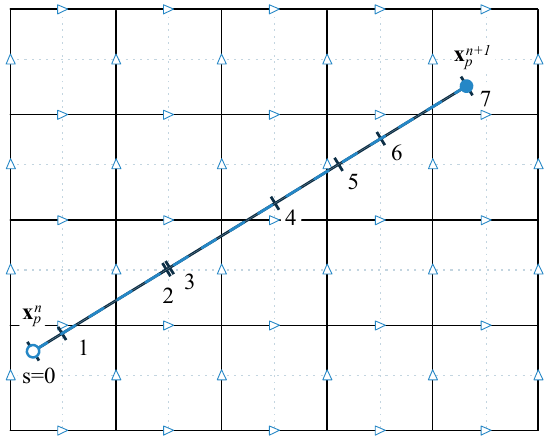}
    \caption{Splitting of the straight trajectory $\xp^n$, $\xp^{n+1}$ into
    segments marked by ticks with index $s$, between the cell surfaces (shown
    by dashed lines). Arrows indicate the components of the electric field
    $E_x$, $E_y$ on the staggered grids: horizontal -- $E_{x;\, i+1/2, j, k}$,
    vertical -- $E_{y;\, i, j+1/2, k}$. The splitting proceeds in a similar way
    for motion along the $z$ direction.}
	\label{fig1}
\end{figure}

Now let us describe the algorithm for interpolating the electric field
$\E_p^{n+1/2}$ from grid to particle. A naive implementation of field
interpolation between $\xp^n$ and $\xp^{n+1}$ does not guarantee charge
conservation on the grid, since a particle may cross cell boundaries, and the
contribution to the current would be ``smeared'' incorrectly. To avoid this, we
follow the idea of Ref.~\cite{chen2023} and split the trajectory into segments,
each of which lies within a single cell (Fig.~\ref{fig1}). For each segment
$s$, its own contribution to the grid nodes is computed, after which these
contributions are averaged with the weight $\Delta x^s_p / \Delta x^n_p$
($\Delta \xp^s = \xp^{s+1} - \xp^s$). In other words, the current is collected
not from the final positions of the particle, but along its entire trajectory
during the step, which ensures the conservative properties of the scheme when
the current is substituted into Maxwell's equations.

The averaged electric field $\E_p^{n+1/2}$ is computed as
\begin{multline}\label{eq:Epg}
    \E_p^{n+1/2} =
    \sum_g \E_g^{n+1/2} {\bf \bar{S}}({\bf x}_g - \xp^{n+1/2}) \\ = 
    {\bf i} \sum_{i,j,k} E_{x;\, i+1/2, j, k}^{n+1/2} S_{xx} +
    {\bf j} \sum_{i,j,k} E_{y;\, i, j+1/2, k}^{n+1/2} S_{yy} +
    {\bf k} \sum_{i,j,k} E_{z;\, i, j, k+1/2}^{n+1/2} S_{zz}.
\end{multline}
Here the interpolation functions $S_{xx}$, $S_{yy}$, $S_{zz}$ are structurally
analogous to those used in the Esirkepov current decomposition
\cite{esirkepov2001}, but they are computed on each segment $s$ ($\Delta \xp^s
= \xp^{s+1} - \xp^s$) and then averaged over the entire length of the traversed
path $\Delta \xp^n = \xp^{n+1} - \xp^n$:
\begin{align}
    S_{xx} = \sum_{s \in n} \frac{\Delta x_p^s}{\Delta x_p^n} \, S_1(x_{i+1/2} -
    x_p^{s+1/2}) \, \frac{1}{3} \bigg[ &S_2(y_j - y_p^{s+1}) S_2(z_k -
    z_p^{s+1}) + \frac{1}{2} \, S_2(y_j - y_p^s) S_2(z_k - z_p^{s+1}) +
    \nonumber \\ \frac{1}{2} \, & S_2(y_j - y_p^{s+1}) S_2(z_k - z_p^s) +
    S_2(y_j - y_p^s) S_2(z_k - z_p^s) \bigg],\\
    S_{yy} = \sum_{s \in n} \frac{\Delta x_p^s}{\Delta x_p^n} \, S_1(y_{i+1/2} -
    y_p^{s+1/2}) \, \frac{1}{3} \bigg[ &S_2(x_j - x_p^{s+1}) S_2(z_k -
    z_p^{s+1}) + \frac{1}{2} \, S_2(x_j - x_p^s) S_2(z_k - z_p^{s+1}) +
    \nonumber \\ \frac{1}{2} \, & S_2(x_j - x_p^{s+1}) S_2(z_k - z_p^s) +
    S_2(x_j - x_p^s) S_2(z_k - z_p^s) \bigg],\\
    S_{zz} = \sum_{s \in n} \frac{\Delta x_p^s}{\Delta x_p^n} \, S_1(z_{i+1/2} -
    z_p^{s+1/2}) \, \frac{1}{3} \bigg[ &S_2(x_j - x_p^{s+1}) S_2(y_k -
    y_p^{s+1}) + \frac{1}{2} \, S_2(x_j - x_p^s) S_2(y_k - y_p^{s+1}) +
    \nonumber \\ \frac{1}{2} \, & S_2(x_j - x_p^{s+1}) S_2(y_k - y_p^s) +
    S_2(x_j - x_p^s) S_2(y_k - y_p^s) \bigg],
\end{align}
where $S_1$ and $S_2$ are the linear and quadratic b-splines. If for the
current decomposition one uses an expression analogous to the interpolation of
the electric field (\ref{eq:Epg})
\begin{equation}
    {\bf J}_g^{n+1/2} = \sum_p q_p \vp^{n+1/2} {\bf \bar{S}} ({\bf x}_g -
    \xp^{n+1/2}),\label{eq:Jn}
\end{equation}
then the laws of conservation of energy and charge are simultaneously satisfied
in difference form. It is important that the same tensor function ${\bf
\bar{S}}$ is used both in the interpolation of the electric field from grid
to particle and in the reverse operation of collecting the current from 
particle to grid. Owing to this symmetry, the change in the kinetic energy
of the particles and the work of the current in Maxwell's equations cancel each
other in the derivation of the discrete energy balance.

\subsection*{Maxwell's equations and self-consistent solution}

The last component of the algorithm is Maxwell's equations in difference form,
describing the evolution of the electromagnetic fields:
\begin{align}
    \left( \E_g^{n+1} - \E_g^{n} \right) / \Delta t &= - {\bf J}_g^{n+1/2} +
    \rot \B_g^{n+1/2},\label{eq:En}\\
    \left( \B_g^{n+1} - \B_g^{n} \right) / \Delta t &= - \rot
    \E_g^{n+1/2}.\label{eq:Bn}
\end{align}
Thus, the transition from the quantities at the previous step $\{ \vp^n, \,
\E_g^n, \, \B_g^n \}$ to the quantities at the new step $n+1$ represents a
joint solution of the nonlinear equations (\ref{eq:vh}, \ref{eq:En},
\ref{eq:Bn}). Maxwell's equations are solved using the Jacobian-Free
Newton--Krylov method (JFNK). The particles
participate in this process not directly, in the form of joint iterations of
the particle velocities and the electromagnetic fields, but through the
contribution of each particle to the half-step current ${\bf J}_g^{n+1/2}$
(\ref{eq:Jn}). The update of the particle velocities occurs separately, at each
outer JFNK iteration. Such a dimension-reduction method was proposed and
described in detail in Refs.~\cite{markidis2011,chen2011,taitano2010}. At each
outer iteration, an approximation for the grid fields is specified, after which
all particles are advanced in these fields, the current ${\bf J}_g^{n+1/2}$ is
collected, and only then the residual of Maxwell's equations is updated. The
nonlinearity of the problem is concentrated in the dependence of the current on
the not-yet-found fields.

Moreover, similarly to the semi-implicit method we
use~\cite{berendeev2024}, the update of the magnetic field from step $n$ to
$n+1$ can be performed separately, outside the joint iteration of
Eqs.~(\ref{eq:En}, \ref{eq:Bn}). For this, the magnetic field $\B_g^{n+1/2}$ is
computed from Eq.~(\ref{eq:Bn}) and substituted into Eq.~(\ref{eq:En}):
\begin{equation}\label{eq:En_mod}
    \frac{ \E_g^{n+1} - \E_g^{n} }{ \Delta t } + \frac { \Delta t }{ 4 } \, \rot \!
    \left( \rot (\E_g^{n+1} + \E_g^n) \right) = - {\bf J}_g^{n+1/2} + \rot
    \B_g^{n}.
\end{equation}
The update of the magnetic field then occurs at the end of the time step, using
Eq.~\ref{eq:Bn}. This form of notation is convenient in that it reduces the
outer nonlinear solver to finding the new electric field, whereas the magnetic
field is reconstructed afterwards explicitly through the discrete Faraday law.
This reduces the dimension of the nonlinear problem compared to the full system
in $\E, \B$ and makes the analysis of the conservation laws presented below
more transparent.

As a result, one step of the algorithm consists of three consistent
operations: advancing the particles in the current approximation of the fields,
collecting the current onto the grid along the segmented trajectory, and
updating the fields from Maxwell's equations. Such a sequence preserves both
the self-consistency of the solution and the discrete integrals of motion.

\section*{Energy and charge conservation}

This section shows that for the described scheme the discrete laws of
conservation of energy and charge are satisfied identically, and not only
approximately. Numerical confirmation of these properties is given in the
section with simulation results. The derivation follows by analogy with
Refs.~\cite{chen2011,chen2023} and relies on the consistency of the field
interpolation and the current decomposition established above.

\subsection*{Discrete energy balance}

Let us begin with the change in the kinetic energy of particles moving according
to Eqs.~(\ref{eq:x}, \ref{eq:v}):
\begin{align}
    \frac{1}{2} \sum_p m_p \left( |\vp^{n+1}|^2 - |\vp^n|^2 \right) &= \sum_p
    m_p \vp^{n+1/2} \left( \vp^{n+1} - \vp^n \right)\\[0.1cm]
    \text{Eq.~(\ref{eq:v})} \: &= \Delta t \sum_p q_p \vp^{n+1/2} \left(
    \E_p^{n+1/2} + \left[ \vp^{n+1/2} \times \B_p^{n+1/2} \right]
    \right)\\[0.1cm]
    \text{$\vp^{n+1/2} \perp (\vp^{n+1/2} \times \B_p^{n+1/2})$} \: &= \Delta t
    \sum_p q_p \vp^{n+1/2} \E_p^{n+1/2}\\[0.1cm]
    \text{Eq.~(\ref{eq:Epg})} \: &= \Delta t \sum_p q_p \vp^{n+1/2} \left(
    \sum_g \E_g^{n+1/2} {\bf \bar{S}}({\bf x}_g - \xp^{n+1/2}) \right)\\[0.1cm]
    \text{permutation of sums} \: &= \Delta t \sum_g \left( \sum_p q_p
    \vp^{n+1/2} {\bf \bar{S}}({\bf x}_g - \xp^{n+1/2}) \right)
    \E_g^{n+1/2}\\[0.1cm]
    \text{Eq.~(\ref{eq:Jn})} \: &= \Delta t \sum_g {\bf J}_g^{n+1/2}
    \E_g^{n+1/2}.
\end{align}
The change in the energy of the electric and magnetic fields is found by
multiplying Eq.~(\ref{eq:En}) by $\E_g^{n+1/2}$ and adding it to
Eq.~(\ref{eq:Bn}) multiplied by $\B_g^{n+1/2}$. Summing over all grid nodes, we
obtain
\begin{multline}
    \frac{1}{2} \sum_g \Big( |\E_g^{n+1}|^2 - |\E_g^{n}|^2 + |\B_g^{n+1}|^2 -
    |\B_g^{n}|^2 \Big)\\
    = \Delta t \sum_g \Big( {\bf J}_g^{n+1/2} \E_g^{n+1/2} + \E_g^{n+1/2} \,
    \rot \B_g^{n+1/2} - \B_g^{n+1/2} \, \rot \E_g^{n+1/2} \Big)\\
    = \Delta t \sum_g \Big( {\bf J}_g^{n+1/2} \E_g^{n+1/2} + \diverg \! \left[
    \E_g^{n+1/2} \times \B_g^{n+1/2} \right] \Big).
\end{multline} 
The last term is the divergence of the Poynting vector and, upon integration
over the volume, turns into a flux through the boundary of the domain. Under
periodic boundary conditions this flux vanishes. Consequently, the change in
the kinetic energy of the particles per step is exactly equal to the change in
the energy of the electromagnetic field with the opposite sign, i.e., the
discrete law of conservation of total energy is satisfied identically. In the
implementation, the accuracy of this balance is limited only by the accuracy of
the nonlinear solver of Maxwell's equations.

\subsection*{Continuity equation}

Let us now show that the discrete continuity equation is satisfied on the grid:
\begin{equation}
    (\rho_g^{n+1} - \rho_g^n) / \Delta t + \diverg {\bf J}_g^{n+1/2} =
    0,\label{eq:dqn}
\end{equation}
where the grid current is determined by the decomposition rule (\ref{eq:Jn}).
Since the continuity equation must hold for each particle separately, it
suffices to verify it for a single macroparticle $p$. Its contribution to the
charge density at step $n$ is given by
\begin{equation}
    \rho_{g, \, p}^n = q_p \, S_2(x_i - x_p^n) S_2(y_j - y_p^n) S_2(z_k -
    z_p^n).
\end{equation}
The particle shape entering the definition of the current is averaged over the
segments $s$ into which the trajectory is split within one time step. Therefore,
for each particle it is convenient to write the continuity equation in the form
\begin{equation}
    \sum_{s \in n} \left[ \left( \rho_{g,\,p}^{s+1} - \rho_{g,\,p}^s \right) /
    \Delta t + \diverg \! \left( q_p \vp^{n+1/2} {\bf \bar{S}} ({\bf x}_g -
    \xp^{s+1/2}) \right) \right] = 0.\label{eq:dqs}
\end{equation}
Identity (\ref{eq:dqs}) is established with the help of a local relation valid
for each segment $s$:
\begin{equation}
    S_2(x_i - x_p^{s+1}) - S_2(x_i - x_p^s) + \frac{\Delta x_p^s}{\Delta x}
    \left[ S_1(x_{i+1/2} - x_p^{s+1/2}) - S_1(x_{i-1/2} - x_p^{s+1/2}) \right] =
    0;
\end{equation}
Substituting this relation into (\ref{eq:dqs}) and summing over all segments,
for particle $p$ we obtain
\begin{multline}
    \sum_{s \in n} \bigg[ S_2(x_i - x_p^{s+1}) S_2(y_j - y_p^{s+1}) S_2(z_k -
    z_p^{s+1}) - S_2(x_i - x_p^s) S_2(y_j - y_p^s) S_2(z_k - z_p^s) \bigg] \\
    = - q_p \bigg\{
        \frac{v^{n+1/2}_{p,\, x}}{\Delta x} \Big[ S_{xx}(x_{i+1/2}, y_j, z_k) -
        S_{xx}(x_{i-1/2}, y_j, z_k) \Big] \\[0.1cm]
        \quad \quad + \frac{v^{n+1/2}_{p,\, y}}{\Delta y} \Big[ S_{yy}(x_i,
        y_{j+1/2}, z_k) - S_{yy}(x_i, y_{j-1/2}, z_k) \Big] \\[0.1cm]
        \quad \quad + \frac{v^{n+1/2}_{p,\, z}}{\Delta z} \Big[ S_{zz}(x_i, y_j,
        z_{k+1/2}) - S_{zz}(x_i, y_j, z_{k-1/2}) \Big] \bigg\} \\[0.1cm]
    = -\diverg \! \left( q_p \vp^{n+1/2} {\bf \bar{S}} ({\bf x}_g - \xp^{s+1/2})
    \right).
\end{multline}
Summing the contributions of all particles, we arrive at
Eq.~(\ref{eq:dqn}). Thus, the discrete Gauss's law, satisfied at the initial
moment, is preserved at all subsequent steps: the change of charge in a cell is
exactly equal to the flux of current through its boundaries.

\section*{Numerical implementation}

The scheme described above is implemented within the family of parallel PIC
codes \texttt{xpic}~\cite{xpic}, available at
\url{https://github.com/vakurshakov/xpic}. The program is written in C++ and
uses the PETSc (Portable Extensible Toolkit for Scientific
Computation)~\cite{petsc2024} library for storing the grid fields, constructing
sparse operators, and solving the resulting algebraic systems. Parallelization
is organized on two levels: decomposition of the computational domain among MPI
processes and distribution of the particle loop among OpenMP threads within
each process. This architecture in \texttt{xpic} is used both in the fully
kinetic electromagnetic model considered here and in its drift-kinetic
extension.

\subsection*{Structure of the time step}

One time step transfers the state of the system from layer $n$ to layer $n+1$.
At the input, the coordinates and velocities of all macroparticles $\xp^n,
\vp^n$ are stored, as well as the grid fields $\E_g^n$ and $\B_g^n$. Since the
fully implicit scheme couples the new fields with the particle trajectories at
the same step, a nonlinear system of equations (\ref{eq:vh}, \ref{eq:Bn},
\ref{eq:En_mod}) arises. It is solved by a nested iterative process: the inner
cycle refines the trajectories of individual particles, while the outer
cycle refines the electromagnetic fields on the grid.

The inner cycle implements the Picard method for the equations of motion
(\ref{eq:x}, \ref{eq:v}) at a fixed approximation of the fields. For each
particle $p$, the unknowns are $\xp^{n+1}$ and $\vp^{n+1}$, whereas
$\vp^{n+1/2}$ is determined from (\ref{eq:vh}). At iteration $k$, the fields
are interpolated to the point $\xp^{n+1/2,k}$ according to the rules
(\ref{eq:Epg}), (\ref{eq:bp}) taking into account the segmentation of the
trajectory; then the coordinate and velocity are updated by the explicit
formulas (\ref{eq:vh}). The residual of the equation of motion with respect to
velocity at iteration $k$ is defined as
\begin{equation}\label{eq:picard_res}
    r_p^{k} = \left\|
        \vp^{n+1,k} - \vp^{n}
        - \Delta t \, \frac{q_p}{m_p}
        \left(
            \E_p^{n+1/2,k} + \left[ \vp^{n+1/2,k} \times \B_p^{n+1/2,k} \right]
        \right)
    \right\|_2,
\end{equation}
The iterations are terminated when the criterion is satisfied
\begin{equation}\label{eq:picard_tol}
    r_p^{k} < \varepsilon_{p}^{a} + \varepsilon_{p}^{r} \, r_p^{0},
\end{equation}
in which $r_p^{0}$ is the residual for the initial approximation at the given
outer iteration. In the implementation, $\varepsilon_{p}^{a} =
\varepsilon_{p}^{r} = 10^{-7}$ and a maximum of 30 Picard iterations is
adopted. In practice, a few iterations are sufficient for convergence, even at
finite values of $q_p B_p \Delta t / m_p$.

In the outer cycle, the root of the nonlinear system of equations for the
fields is sought. After substituting (\ref{eq:Bn}) into (\ref{eq:En}), the
problem reduces to Eq.~(\ref{eq:En_mod}) with respect to the new electric field
$\E_g^{n+1}$, which is conveniently written in the form ${\bf F}(\E_g^{n+1}) =
0$, where
\begin{equation}\label{eq:residual}
    {\bf F}(\E_g^{n+1}) = \frac{\E_g^{n+1} - \E_g^{n}}{\Delta t}
    + \frac{\Delta t}{4} \rot \! \left( \rot (\E_g^{n+1} + \E_g^n) \right)
    + {\bf J}_g^{n+1/2}(\E_g^{n+1}) - \rot \B_g^{n}.
\end{equation}
The residual of this system of equations at iteration $k$ is defined
analogously as $r_f^{k} = \| {\bf F}(\E_g^{n+1,k}) \|_2$. The magnetic field at
the half-step is reconstructed from $\E_g^{n+1}$ by (\ref{eq:Bn}), and the full
field at layer $n+1$ is reconstructed at the end of the step. The nonlinearity
of the residual ${\bf F}$ is entirely concentrated in the dependence of the
current ${\bf J}_g^{n+1/2}$ on the not-yet-found fields through the particle
trajectories. The convergence of the outer solver is controlled by the standard
PETSc~SNES criteria: the iterations are terminated if at least one of the
conditions is satisfied
\begin{align}
    r_f^{k} &< \varepsilon_f^{a}, \nonumber \\[0.1cm]
    r_f^{k} &< \varepsilon_f^{r} \, r_f^{0}, \label{eq:snes} \\[0.1cm]
    \|\E_g^{n+1,k} - \E_g^{n+1,k-1}\|_2 &< \varepsilon_f^{s} \| \, \E_g^{n+1,k}\|_2, \nonumber
\end{align}
where $r_f^{0} = \|{\bf F}(\E_g^{n+1,0})\|_2$ is the residual of the equation
for the initial approximation (in the implementation, $\E_g^{n}$ is taken). In
the simulations, $\varepsilon_f^{a} = \varepsilon_f^{r} = \varepsilon_f^{s} =
10^{-7}$ is used with a maximum of 1000 nonlinear iterations.

\subsection*{JFNK method with particle suppression}

Direct application of the Newton--Krylov method (JFNK) to the full
field--particle system is inefficient: the number of unknowns associated
with particles usually exceeds the number of grid nodes by several orders of
magnitude, and the construction of Krylov subspaces for such a system requires
unacceptable amounts of memory~\cite{angus2023,chen2011,markidis2011}.
Therefore, following the particle enslavement / particle-suppressed JFNK
(PS-JFNK) approach~\cite{angus2023,chen2011,markidis2011,taitano2010}, we
exclude the particle variables from the unknown vector of the outer solver.

The idea consists in nonlinear elimination: for a given trial $\E_g^{n+1}$,
Eqs.~(\ref{eq:vh}) are solved independently for each particle until
(\ref{eq:picard_tol}) is satisfied, after which the current is collected by
(\ref{eq:Jn}) and the residual (\ref{eq:residual}) is computed. The particles
do not enter the Jacobian-free solver directly and participate in it only
through the current ${\bf J}_g^{n+1/2}(\E_g^{n+1})$. The outer JFNK, implemented
by means of PETSc, requests at each nonlinear iteration only the value of
${\bf F}$ and, if necessary, its derivative along a trial vector
\begin{equation}
    \frac{\partial {\bf F}}{\partial \E} \cdot {\bf V} \approx
    \frac{{\bf F}(\E_g^{n+1} + \varepsilon {\bf V}) - {\bf F}(\E_g^{n+1})}{\varepsilon},
\end{equation}
which makes it possible to avoid the explicit construction of the Jacobian
matrix. The linear systems for the Newton corrections are solved by the GMRES
method; the outer iterations continue until any of the criteria (\ref{eq:snes})
is satisfied. For each call of ${\bf F}$, it is necessary to traverse all
particles with inner Picard iterations, so the main computational load falls on
the current computation.

A special feature of this architecture is that the outer solver does not access
the particles directly: at each of its iterations it only specifies an
approximation for $\E_g^{n+1}$, while the current ${\bf J}_g^{n+1/2}$ is formed
by the inner advancement of the macroparticles in these fields. The
self-consistency of the fields and particles is achieved through this pair of
operations. In practice, to accelerate convergence, a warm start is used: the
inner Picard cycle for a new field approximation begins not from $\xp^n,
\vp^n$, but from the already found $\xp^{n+1}, \vp^{n+1}$ at the previous outer
iteration.

\subsection*{Parallel organization of computations}

The decomposition of the computational domain among MPI processes is inherited
from the PETSc grid structures: each process owns its block of the Yee grid and
the macroparticles lying in it. The operators of curl, divergence, and
gradients on the grid are formed once as sparse matrices and then multiplied by
the grid vectors of the fields. The particles are sorted by cells; during a
step, a macroparticle may leave the subdomain of a process, so an exchange with
neighbors is performed after the coordinates are updated. Periodic boundaries
are taken into account before splitting the trajectory into segments.

Within an MPI process, the loop of particle advancement and current collection
is distributed among OpenMP threads over cells. The contribution of each
particle to the grid nodes is added atomically, which excludes contention for
memory access when several threads write simultaneously to common nodes. The
segmentation of the trajectory when flying through several cells and the
consistent interpolation of the fields are implemented uniformly for all
transfer operations between particles and the grid, which is necessary for the
exact fulfillment of the difference conservation laws proved above.

The full step of the implicit algorithm includes: (i)~solving
(\ref{eq:En_mod}) by the PS-JFNK method with inner particle advancement;
(ii)~reconstruction of $\B_g^{n+1}$ by (\ref{eq:Bn}); (iii)~migration of
particles between cells and MPI subdomains. The numerical experiments of the
next section were performed in this implementation.

\section*{Simulation results}

The correctness of the implemented scheme is verified on three problems:
equilibrium Maxwellian plasma at $\Delta x \gg \lambda_D$, the linear stage of
the Weibel instability, and the formation of a layer with $\beta \sim 1$ under
continuous plasma injection into a vacuum magnetic field. In all tests, the
standard explicit scheme~\cite{kurshakov2023} (explicit), the semi-implicit
ECSIM variants without current correction (ecsim) and with correction
(ecsimcorr)~\cite{berendeev2024}, and the fully implicit scheme described above
(eccfim) are compared.

\subsection*{Equilibrium plasma}

\begin{figure}[htb!]
    \centering
	\includegraphics[scale=0.7]{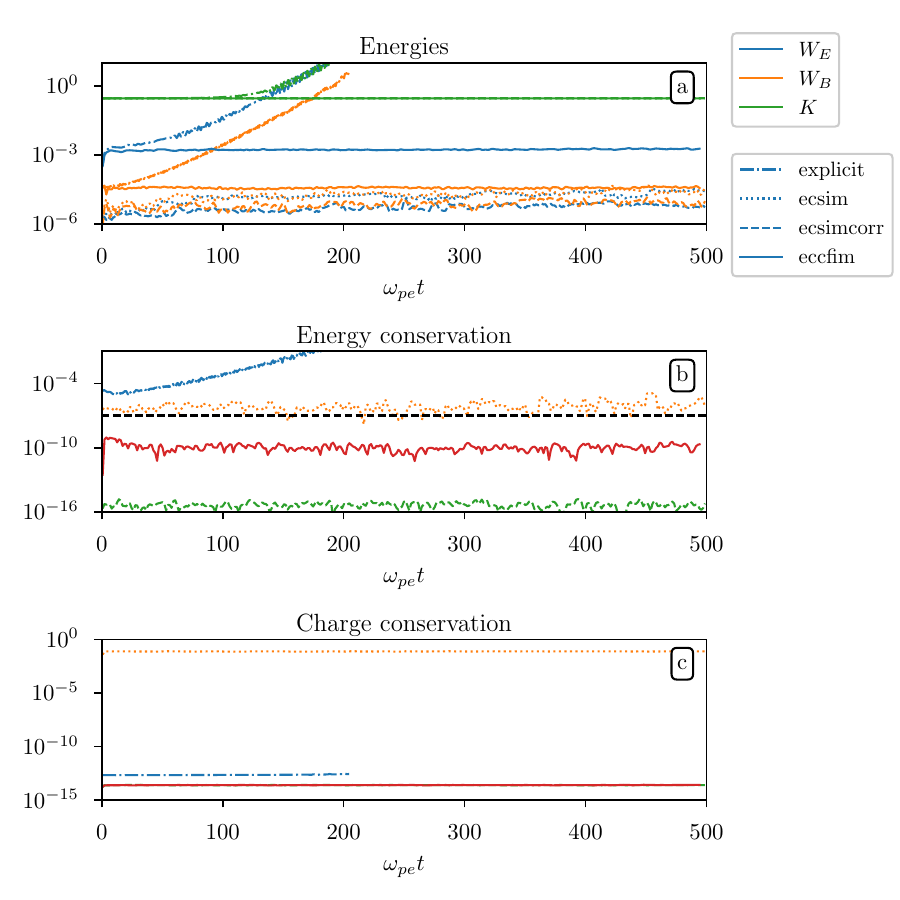}
    \caption{Equilibrium Maxwellian plasma at $\Delta x \gg \lambda_D$. Lines:
    dash-dotted~--- explicit, dots~--- ecsim, dashed~--- ecsimcorr, solid~---
    eccfim. (a)~Energies of the electric field $W_E$ (blue), magnetic field
    $W_B$ (orange), and kinetic energy of particles $K$ (green).
    (b)~Relative change of the total energy $\Delta(W_E + W_B + K)$; black
    dashed line~--- solver tolerance level $\varepsilon = 10^{-7}$.
    (c)~Residual of the discrete continuity equation
    $\bigl[\sum_g (\Delta\rho_g/\Delta t + \diverg {\bf J}_g)^2\bigr]^{1/2}$.}
	\label{fig2}
\end{figure}
The first test reproduces the setup of Ref.~\cite{berendeev2024}. A uniform
Maxwellian plasma in a cubic domain with periodic boundaries is considered. The
ions are immobile, the electron temperature is $T_e = 100$~eV ($v_T = 0.014$ in
the adopted units). The grid step $\Delta x = \Delta y = \Delta z = 0.3$ exceeds
the Debye radius by approximately a factor of 20 ($\lambda_D / \Delta x =
0.046$), so in explicit schemes a numerical growth of energy is expected due to
the nonlinear interaction of grid modes~\cite[\S 8-13]{birdsall2018}. The time
step $\Delta t = 5\Delta x = 1.5$ exceeds the CFL condition for the explicit
scheme, $\Delta t_{\mathrm{CFL}} = \Delta x/\sqrt{3} \approx 0.17$; for the
explicit run a smaller step $\Delta t = 0.15$ is used in order to separately
demonstrate the development of the same grid instability. The domain size is
$N_x = N_y = N_z = 10$, and the number of macroparticles per cell is $N_p =
50$.

The results are shown in Fig.~\ref{fig2}. In the explicit scheme the energy
grows rapidly (Fig.~\ref{fig2}a), whereas all implicit variants keep the total
energy at the level of the solver error $\varepsilon = 10^{-7}$ without any
tendency to grow (Fig.~\ref{fig2}b). For ecsim and ecsimcorr this level is set
by the accuracy of the linear field system, while for eccfim it is set by the
accuracy of the nonlinear solver (\ref{eq:snes}). The continuity equation is
satisfied for explicit, ecsimcorr, and eccfim; in ecsim the local Gauss's law
is not satisfied by construction of the scheme (Fig.~\ref{fig2}c). Thus, on a
coarse grid and at a step $\Delta t \gg \omega_{pe}^{-1}$, the fully implicit
scheme conserves both discrete integrals of motion.

\subsection*{Weibel instability}

\begin{figure}[htb!]
    \centering
	\includegraphics[width=\linewidth]{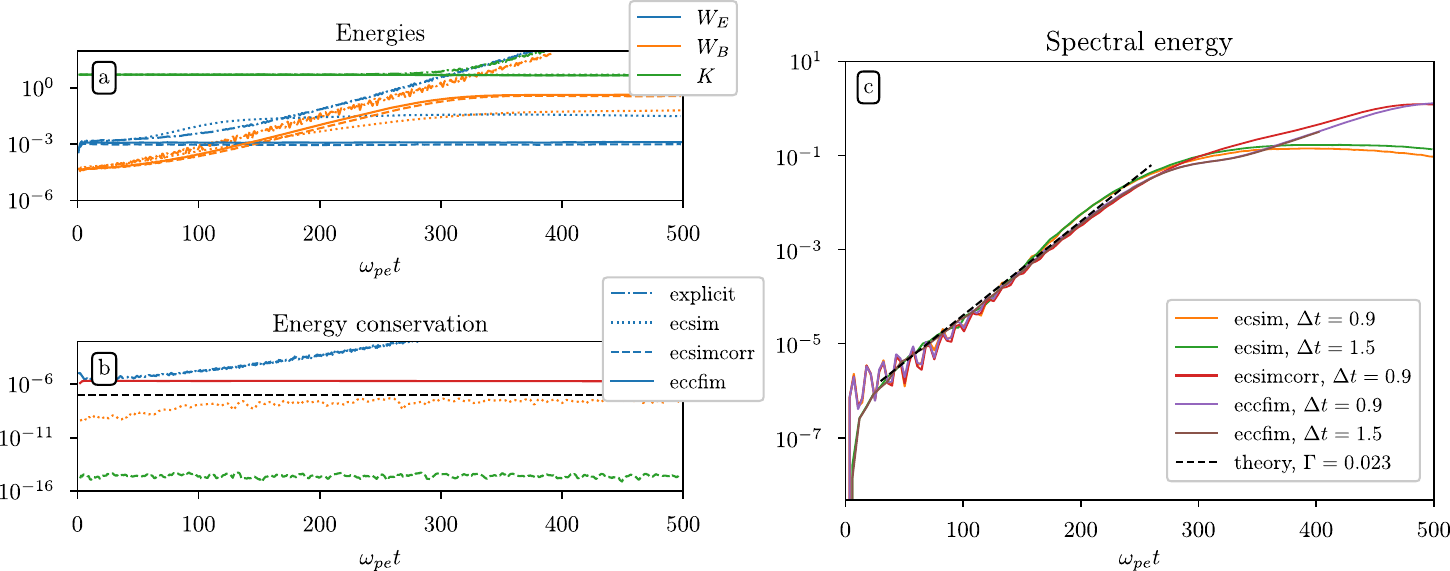}
    \caption{Weibel instability. (a)~Energies $W_E$, $W_B$, and $K$;
    (b)~conservation of the total energy $\Delta(W_E + W_B + K)$; the
    designations of the curves are the same as in Fig.~\ref{fig2}.
    (c)~Energy of the most unstable mode $B_x(k_x=1,\,k_y=1)$ for the implicit
    schemes; dashed line~--- theoretical growth rate
    $\Gamma_{\max} \approx 0.023$.}
	\label{fig3}
\end{figure}
The second test verifies the ability of the scheme to describe low-frequency
electromagnetic dynamics at $\omega \ll \omega_{pe}$. A uniform plasma with an
anisotropic electron temperature $T_\| = 1$~keV, $T_\perp =
0.1$~eV~\cite{berendeev2024,weibel1959} is specified. The steps $\Delta x$ and
$\Delta t$ are the same as in the previous test; the domain size is
$N_x = N_y = 30$, $N_z = 5$.

As in the equilibrium case, the explicit scheme gives a nonphysical growth of
energy (Fig.~\ref{fig3}a). The implicit schemes are stable at
$\omega_{pe}\Delta t = 1.5$, i.e., at a step exceeding the plasma period but
still resolving the instability growth rate ($\Gamma_{\max}\Delta t \approx
0.035$). The total energy is conserved to within the solver accuracy
(Fig.~\ref{fig3}b), and the continuity equation is satisfied for the schemes
with consistent current decomposition. In Fig.~\ref{fig3}c the energy of the
mode $B_x(k_x=1,\,k_y=1)$ for ecsim, ecsimcorr, and eccfim grows with the same
growth rate, close to the theoretical value $\Gamma_{\max} \approx 0.023$.
At the stage of nonlinear saturation, however, one can see some difference: ecsimcorr and
eccfim continue to coincide with each other, while ecsim underestimates the
amplitude of the dominant unstable mode, since a portion of the non-equilibrium
energy is expended on the excitation of enhanced electromagnetic noise.
This confirms that charge conservation is an important ingredient in energy-conserving
PIC models and that our fully implicit discretization does not distort
the dynamics of this instability.

\subsection*{Formation of a layer with $\beta \sim 1$}

The third test is considered as an analogue of the target regime of an open
trap: the formation of a one-dimensional plasma layer with high $\beta$ under
continuous injection into an initially uniform vacuum field $\mathbf{B} =
(0,0,B_v)$. The setup coincides with
Refs.~\cite{kurshakov2023,berendeev2024} and is related to the experiment at
the CAT facility~\cite{bagryansky2016}, where a high-pressure plasmoid is
created by powerful injection of neutral atoms into a target plasma.

\begin{figure}[htb!]
    \centering
	\includegraphics[width=0.6\linewidth]{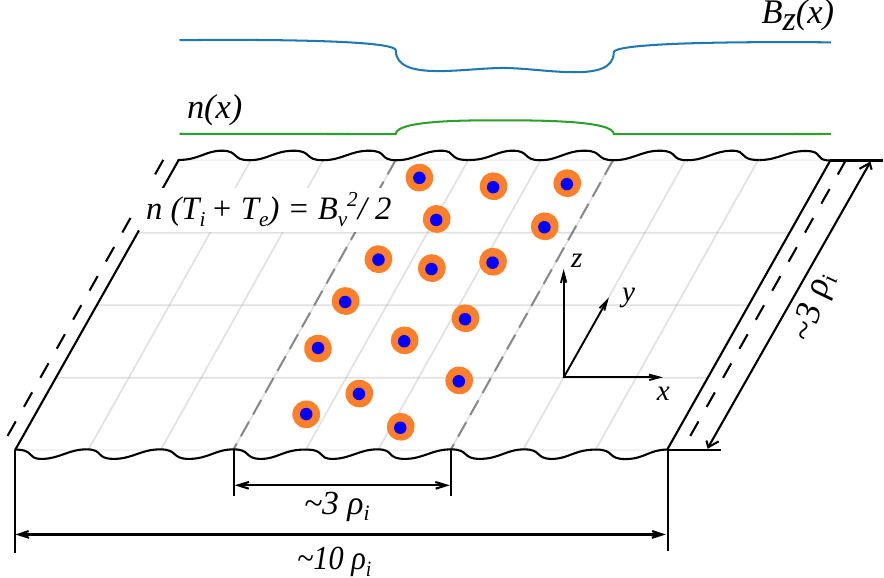}
    \caption{Schematic of the computational domain under continuous plasma
    injection into a vacuum magnetic field: the central injection zone and
    absorbing layers along the $x$ axis.}
	\label{fig4}
\end{figure}
Electron--ion pairs with temperatures $T_i = 10$~keV, $T_e = 2$~keV, and a mass
ratio $m_i/m_e = 16$ are uniformly injected into the central injection zone
(Fig.~\ref{fig4}) so that by the time $\tau = 1000\,\omega_{pe}^{-1}$ the
density linearly reaches $n_0 = 10^{13}$~cm$^{-3}$ (this value is taken as
unity). The vacuum field is $B_v = \sqrt{2 T_i/(m_e c^2)} \approx 0.1978$; by
$t \approx \tau$ one expects $\beta \approx 1$. Along the field the domain is
truncated to $L_z = 5\Delta z$; the steps are $\Delta t = 0.5$ and $\Delta x =
\Delta y = \Delta z = 1.0$. There are 200 macroparticles per cell at unit
density. The boundaries are periodic in $y$ and $z$, and absorbing in $x$ for
both the fields and the particles.

\begin{figure}[htb!]
    \centering
	\includegraphics[width=0.9\linewidth]{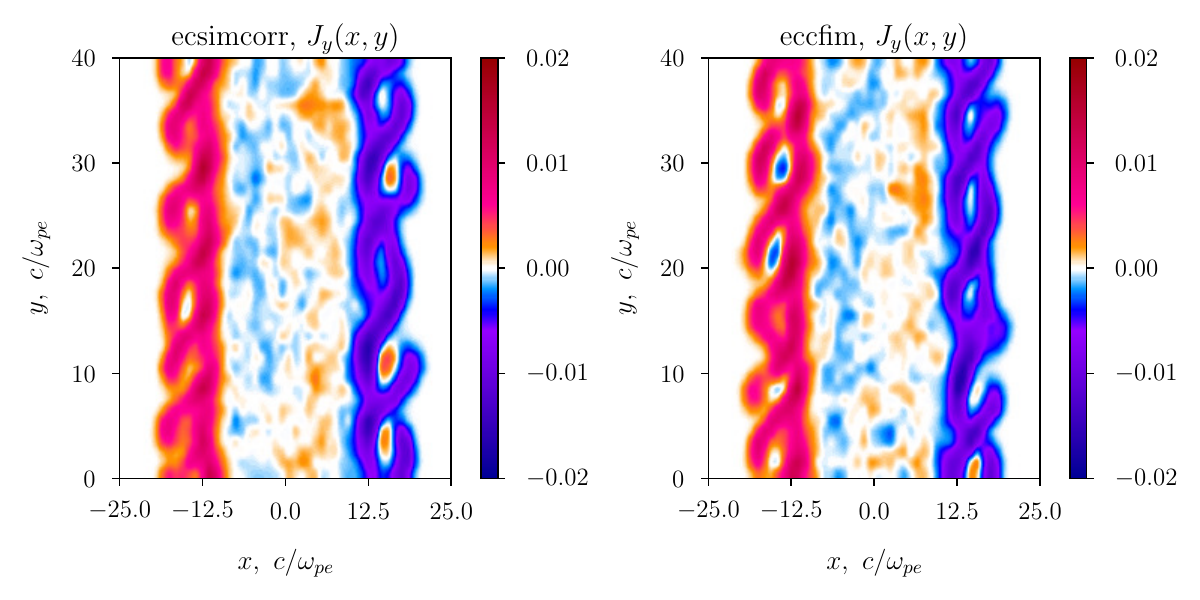}
    \caption{Current density $J_y(x,y)$ at $t = 0.6\tau$ in the injection
    problem: left~--- semi-implicit scheme ecsimcorr, right~--- fully implicit
    scheme eccfim. The same color scale is used in both panels.}
	\label{fig5}
\end{figure}

\begin{figure}[htb!]
    \centering
	\includegraphics[width=\linewidth]{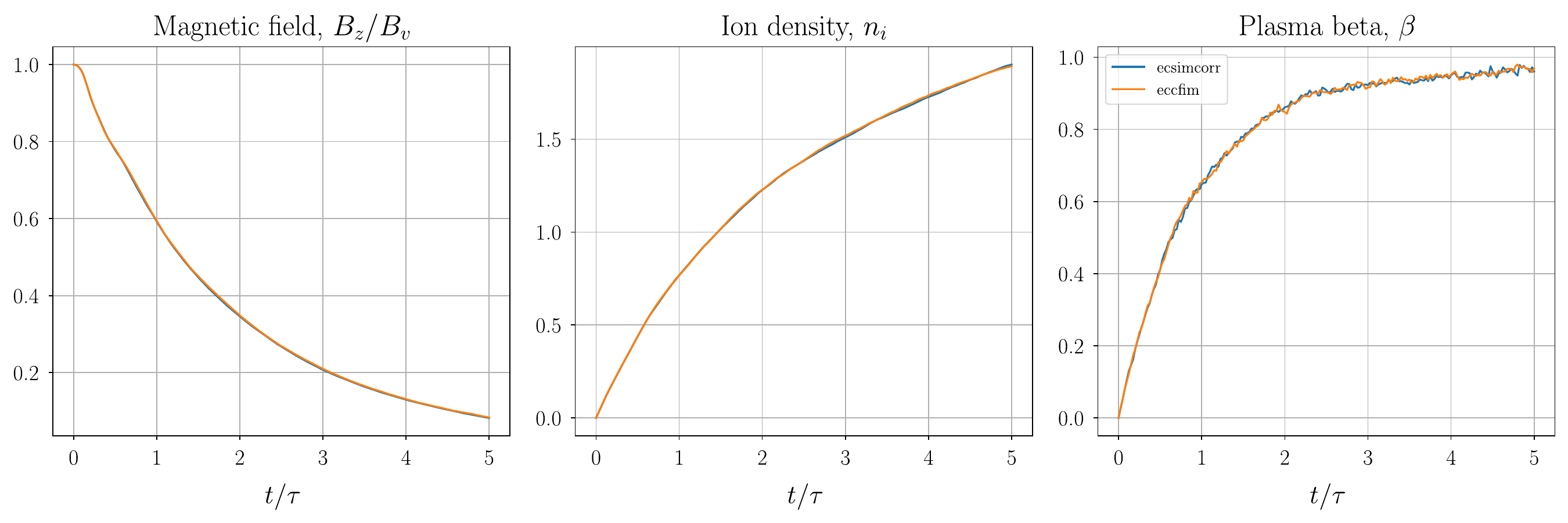}
    \caption{Dynamics at the center of the injection zone: the ratio $B_z/B_v$,
    the ion density $n_i$, and the parameter $\beta$ for the semi-implicit
    (ecsimcorr) and fully implicit (eccfim) schemes.}
	\label{fig6}
\end{figure}

Both schemes reproduce the drift-ion-cyclotron instability at the boundaries of
the injection zone. Figure~\ref{fig5} shows the current density $J_y(x,y)$ at
$t = 0.6\tau$ for ecsimcorr (left) and eccfim (right). The vortex-like
structures forming at the edges of the injection zone have the same spatial
scale and qualitatively the same pattern in both schemes, which confirms that
the fully implicit discretization reproduces the spatial structure of the
instability.
Figure~\ref{fig6} compares the dynamics at the center of the injection zone.
The curves $B_z/B_v$, $n_i$, and $\beta$ for ecsimcorr and eccfim practically
coincide.
The agreement in the macroscopic profiles shows that the transition to a
fully implicit discretization does not change the physics of the transition
layer formation at the chosen parameters.

\begin{figure}[htb!]
    \centering
	\includegraphics[width=0.7\linewidth]{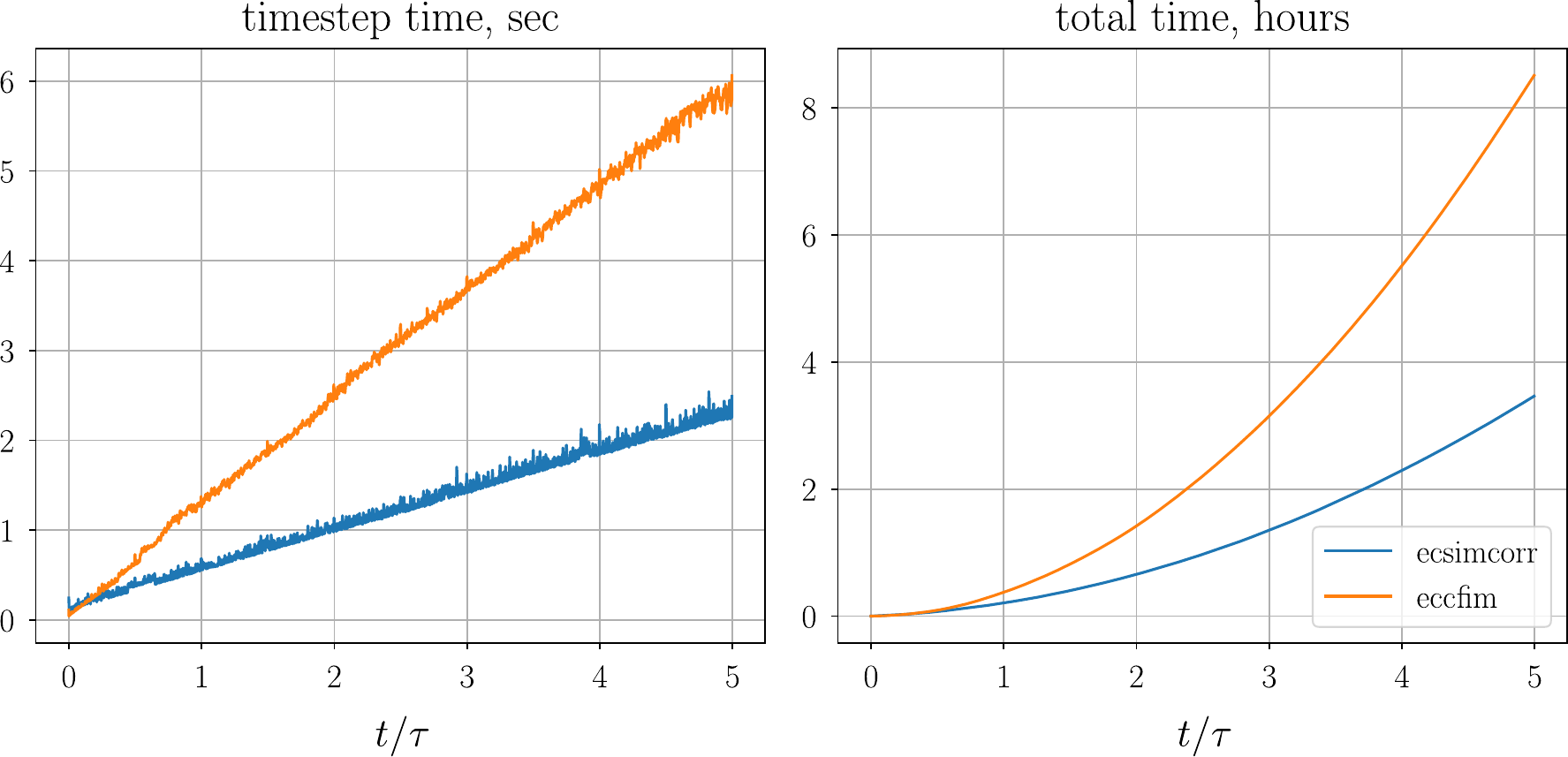}
    \caption{Performance on the injection problem: on the left~--- time of one
    step, on the right~--- integral computation time up to $t = 5\tau$ for
    ecsimcorr (blue) and eccfim (orange).}
	\label{fig7}
\end{figure}

According to Fig.~\ref{fig7}, at these steps ($\Delta x \approx \Delta t$)
eccfim is about 2.5 times slower than ecsimcorr, both in the time of one step
and in the integral time up to $t = 5\tau$. This is expected: each outer
evaluation of the residual (\ref{eq:residual}) requires a full pass over the
particles with Picard iterations. Therefore, below we separately investigate
at which combinations of $\Delta x$ and $\Delta t$ the fully implicit scheme
can become comparable to the semi-implicit one in performance.

\section*{Comparison of the efficiency of PIC methods}

\begin{figure}[htb!]
    \centering
	\includegraphics[width=\linewidth]{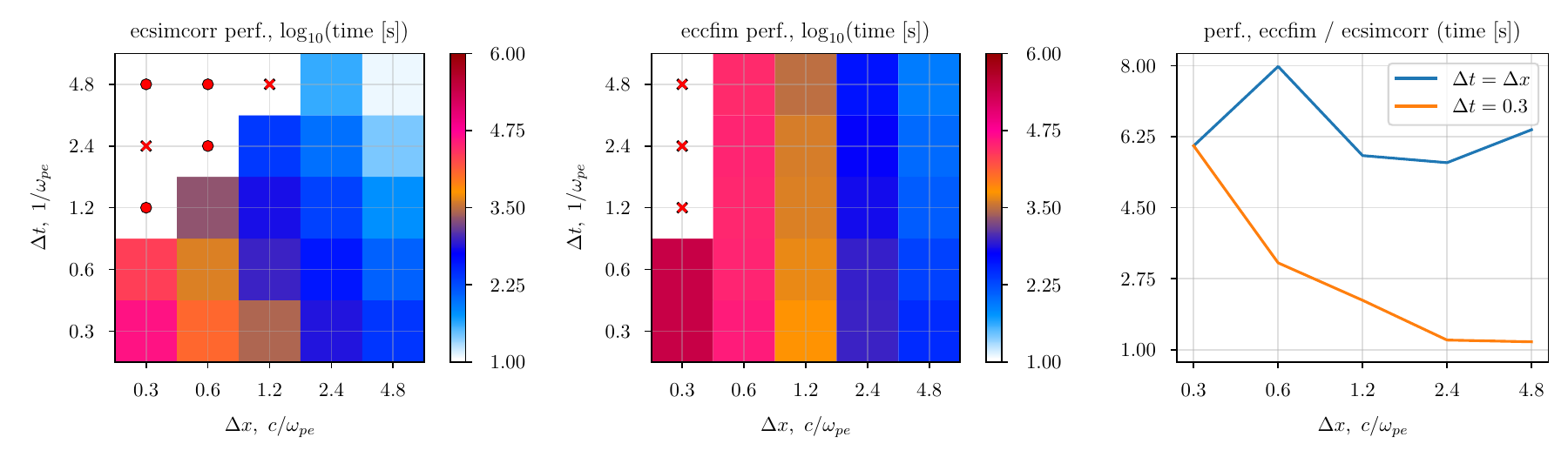}
    \vspace*{-0.7cm}
    \caption{Performance map. Left: $\log_{10}(t\,[\mathrm{s}])$ for ecsimcorr.
    Center: $\log_{10}(t\,[\mathrm{s}])$ for eccfim. Right: the ratio of the
    computation times, $t_{\mathrm{eccfim}}/t_{\mathrm{ecsimcorr}}$, as a function of $\Delta x$ for
    two families of runs: $\Delta t = \Delta x$ (blue) and $\Delta t = 0.3$
    (orange). Crosses and circles mark runs that terminated abnormally.}
	\label{fig8}
\end{figure}
For a quantitative assessment of the computational efficiency of ecsimcorr and
eccfim, a series of simulations was carried out on the same physical problem:
Maxwellian electrons with $T_e = 30$~eV, domain $L_x = L_y = 60\,c/\omega_{pe}$,
$N_z = 5$ cells, physical time $T = 300\,\omega_{pe}^{-1}$, and $N_p = 1000$
particles per cell. The steps were varied in the range
$$
    \Delta x\,\omega_{pe}/c,\;
    \Delta t\,\omega_{pe}
    \in \{0.3,\; 0.6,\; 1.2,\; 2.4,\; 4.8\}.
$$
The measure used is the real computation time $t\,[\mathrm{s}]$. In
Fig.~\ref{fig8}, the left panel shows $\log_{10} t$ for ecsimcorr, the center
panel shows $\log_{10} t$ for eccfim, and the right panel shows the ratio of
the computation times on the same grid, $t_{\mathrm{eccfim}}/t_{\mathrm{ecsimcorr}}$.
Crosses and circles mark runs that terminated abnormally (numerical instability
or solver divergence).

The right panel shows the ratio $t_{\mathrm{eccfim}}/t_{\mathrm{ecsimcorr}}$ for
two families of runs: $\Delta t = \Delta x$ (blue) and $\Delta t = 0.3$ (orange).
When both steps are increased together ($\Delta t = \Delta x$), 
the ratio remains roughly constant at 6--8, i.e., the fully implicit scheme is
consistently more expensive. However, when the time step is fixed ($\Delta t = 0.3$)
and only the spatial step is increased, the ratio decreases monotonically from
about 6 at $\Delta x = 0.3$ to unity at $\Delta x \geq 2.4$. Thus, it is the
increase of the spatial step at a fixed time step, rather
than the simultaneous increase of both, that makes the fully implicit scheme
competitive. This behavior is explained by the growing cost of assembling
the Lapenta mass matrix in ecsimcorr as the total number of particles and
the number of cell crossings per step increase, whereas in eccfim the cost per
step grows more slowly with $\Delta x$. Thus, at large spatial steps the fully
implicit approach can have a competitive performance compared to the semi-implicit
one, which is promising for long-term three-dimensional simulations of
high-$\beta$ plasma.

\section*{Conclusion}

This work implements a fully implicit electromagnetic PIC method for
simulating high-$\beta$ plasma typical to open magnetic traps. The method makes it
possible not to resolve the electron plasma frequency and the Debye radius
while exactly satisfying the discrete laws of conservation of energy and
charge. The algorithm is based on the joint solution of the particle equations
of motion and Maxwell's equations with consistent interpolation of the electric
field along the trajectory and current collection by the Esirkepov method; the
nonlinear system is solved by nested Picard and PS-JFNK processes.

The test simulations confirm the stability and conservativeness of the scheme.
In equilibrium plasma at $\Delta x \gg \lambda_D$, the explicit scheme gives a
nonphysical growth of energy, whereas ecsim, ecsimcorr, and eccfim keep the
total energy at the level of the solver accuracy $\varepsilon = 10^{-7}$; for
the schemes with consistent current decomposition, the continuity equation is
also satisfied. At the linear stage of the Weibel instability, the implicit
schemes reproduce the theoretical growth rate at $\omega_{pe}\Delta t = 1.5$,
i.e., in a regime inaccessible to explicit methods because of the CFL
constraint. In the problem of the formation of a layer with $\beta \sim 1$,
the dynamics of $B_z$, $n_i$, and $\beta$ for eccfim coincides with ecsimcorr,
including the development of the drift-ion-cyclotron instability at the
boundaries of the injection zone.

In terms of performance, when both steps are increased together
($\Delta t = \Delta x$), eccfim is approximately six to eight times more
expensive than ecsimcorr, as follows from the ratio
$t_{\mathrm{eccfim}}/t_{\mathrm{ecsimcorr}}$. However, when the time step is
fixed and only the spatial step is increased, this ratio decreases and, for
$\Delta x \geq 2.4$, the computation times become comparable. This makes the
fully implicit scheme promising for long-term three-dimensional simulations of
high-$\beta$ plasma.

\vspace{0.5cm}

{\bf CRediT authorship contribution statement}

{\bf Vladislav A. Kurshakov}: Investigation, Software, Validation,
Visualization, Writing -- original draft.

{\bf Igor V. Timofeev}: Conceptualization, Investigation, Supervision,
Validation, Writing -- original draft.

\vspace{0.5cm}
{\bf Declaration of competing interest}

The authors declare that they have no known competing financial interests or
personal relationships that could have appeared to influence the work reported
in this paper.

\bibliographystyle{elsarticle-num}
\bibliography{kurshakov2026a}

\end{document}